\documentclass[pre,twocolumn,showpacs]{revtex4-1}
\usepackage{graphicx}
\usepackage{amsmath,amssymb}

\usepackage{color}

\newcommand{\me}{{\rm e}}

\begin{document}

\title{Sequence selection in an autocatalytic binary polymer model}

\author{Shinpei Tanaka$^1$, Harold Fellermann$^2$, and Steen Rasmussen$^{2,3}$}
\affiliation{$^1$Graduate School of Integrated Arts and Sciences, Hiroshima
University, 1-7-1 Kagamiyama, Higashi-Hiroshima 739-8521, Japan \\
$^2$Center for Fundamental Living Technology (FLinT)
Department of Physics, Chemistry and Pharmacy
University of Southern Denmark, Campusvej 55 5230 Odense M, Denmark \\
$^3$Santa Fe Institute, 1399 Hyde Park Rd, Santa Fe NM 87501, USA}

\begin{abstract}
An autocatalytic pattern matching polymer system is studied as an abstract model for chemical ecosystem evolution. 
Highly ordered populations with particular sequence patterns appear spontaneously out of a vast number of possible states. 
The interplay between the selected microscopic sequence patterns and the macroscopic cooperative structures is examined. 
Stability, fluctuations, and evolutionary selection mechanisms are investigated for the involved self-organizing processes.
\end{abstract}

\maketitle
The emergence of autocatalytic structures in model chemistries has been a prominent research subject in complex systems, artificial life and origins of life studies. 
Earlier work is mainly concerned with the emergence and dynamics of autocatalytic sets of random cross-catalytic molecules \cite{Oparin1961,Eigen1971,Kauffman1971,Rasmussen1985,Farmer1986,Kauffman1986,Rasmussen1989,
Bagley1991,Hordijk2011}, 
or focused on the detailed dynamics of template directed catalytic ligation~\cite{Kiedrowski1986,Fernando2007,Fellermann2011,Obermayer2011,Walker2012}.
In this study we investigate how the detailed sequence patterns of self-replicating polymers impact the frequency and stability of emerging population structures.
We study a system of binary polymers, where each polymer can replicate itself by exact ligation of two matching subsequences.
We demonstrate how the microscopic interactions resulting from polymer sequence details dictate selection, population size as well as the frequency and the stability of the evolving, macroscopic, cooperative structures.  
These findings could have implications for early earth information
polymers, the practical design of protocell information polymer networks
(information-metabolisms) as well as more general issues concerning the
emergence of cooperation in complex systems.

Our binary polymers are modeled as strings of monomers over the alphabet $\mathcal A = \left\{0,1\right\}$.
We denote the set of polymers by $\mathcal A^*$ and we write $|k|$ to express the length of polymer $k\in\mathcal A^*$.
In the simplest realization for interaction between polymers, three types of reactions are introduced:
decomposition of a strand into any two substrands with rate $c_0$, random ligation of two strands with rate $c_1$, and autocatalytic ligation with rate $c_2$. Formally:
\begin{align}
	\label{eq_decomposition}
	l.m  &\xrightarrow{c_0} l+m \\
	\label{eq_uncatalyzed}
	l+m &\xrightarrow{c_1} l.m \\
	\label{eq_catalyzed}
	l+m+l.m &\xrightarrow{c_2} 2 \; l.m 
\end{align}
for all $l,m \in \mathcal A^*$, where $l.m$ represents the concatenation of strands $l$ and $m$.
Here, we consider only closed systems where material is conserved.  
However, an energy flow is implicitly assumed for the autocatalytic reactions.

Populations are represented by a multidimensional vector $x$ where each species $k\in \mathcal A^*$ defines one dimension and its concentration $x_k \in \mathbb{R^+}$ specifies a coordinate.

\begin{figure}
 \begin{center}
  \includegraphics[width=.77\columnwidth]{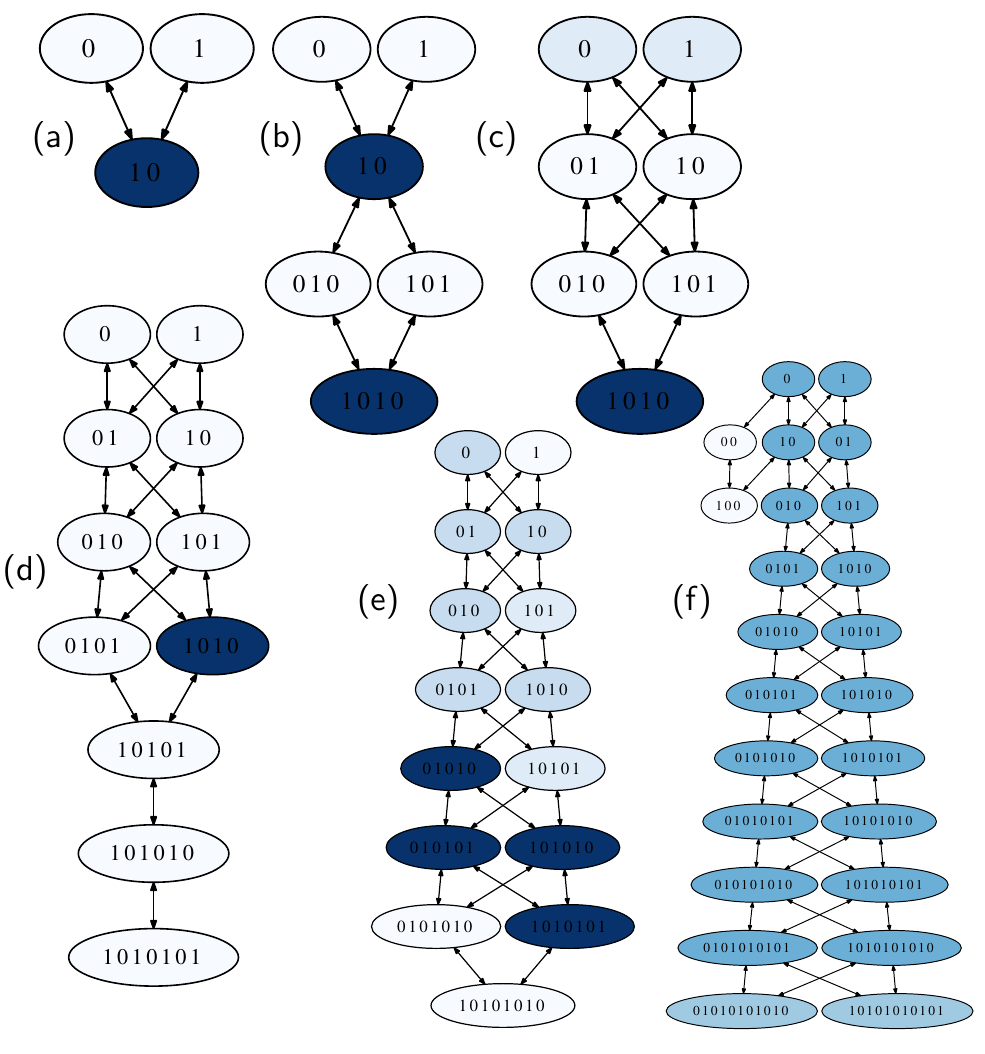} 
  \caption{A stabilization process of ``bootlace'' structure. The darker
  the color the larger the population. (a) $t=1.1$ (b) $t=1.2$ (c)
  $t=1.3$ (d) $t=2.1$ (e) $t=4.8$ (f) $t=230$. Only edges connecting species in two adjacent layers are drawn.} \label{process}
 \end{center}
\end{figure}

Assuming that the law of mass action holds in this system, the reaction kinetic equations can be written in terms of the molar concentrations $x_k$ of all species $k$ as
\begin{align}
	\frac{dx_k}{dt} &= c_0\left(\sum_{\substack{k.j=i \\ j.k=i}}x_i-\sum_{i.j=k}x_k\right) \nonumber \allowdisplaybreaks[1] \\
	&+ c_1\left(\sum_{i.j=k}x_ix_j-\sum_{\substack{k.j=i \\ j.k=i}}x_jx_k\right) \nonumber \allowdisplaybreaks[1] \\
	&+ c_2\left(\sum_{i.j=k}x_ix_jx_k-\sum_{\substack{k.j=i \\ j.k=i}}x_ix_jx_k\right)\equiv f_k.
	\label{eq1}
\end{align}
where $i,j,k \in \mathcal A^*$. Summations $\sum_{i.j=k}$ and $\sum_{k.j=i, j.k=i}$ consider every pair of polymers $i$ and $j$ which satisfy $i.j=k$, and $k.j=i$ or $j.k=i$, respectively. 
Note that the latter involves an infinite number of summands. 

In the limit of purely random ligation ($c_1>0$, $c_2=0$), there is an exponential stationary solution that satisfies $f(x^*)=0$:
\begin{equation}
	x^*_k = (c_0/c_1)\me^{-b|k|},
	\label{exponential}
\end{equation}
where $b$ is a constant determined by the boundary condition.

In the presence of autocatalysis, a species can maintain itself in sufficient concentration, thereby creating via decomposition the set of its substrings.
We call this set a ``clan'' and we call the longest species of a clan its ``chief''.
Two or more chiefs can coexist as seen, for example, in Fig.~\ref{process}(f), where 01010101010 and 10101010101 coexist and share most of their clan members while 100 is also a chief and shares its members of 01, 10, 0, and 1 with the above two chiefs.

In the purely autocatalytic limit ($c_1=0$, $c_2>0$), the exponential solution \eqref{exponential} cannot satisfy $f(x^*)=0$.  
Instead, there is a somewhat counter-intuitive constant solution
\begin{align}
	x^*_k&=
	\begin{cases}
	 \sqrt{c_0/c_2}\equiv \bar{x}, & k\in
	 \mathcal{A}^\dagger\backslash \partial\mathcal{A}^\dagger \\
	 0, & k \not\in \mathcal{A}^\dagger,
	\end{cases}
	\label{constant}
\end{align}
where $\mathcal{A}^\dagger$ is the union of all clans and $\partial\mathcal{A}^\dagger$ the set of all chiefs, both given by the initial condition.
Interestingly, chief populations $x^*_m$ are unconstrained in this case: the stationarity condition $f_m=0$ demands that
\begin{align}
	x^*_m\left(-c_0\sum_{i.j=m}1+c_2\sum_{i.j=m}x^*_ix^*_j\right)&=0.
 \label{eq_chief}
\end{align}
Since $x^*_m\neq 0$, $\sum_{i.j=m}x^*_ix^*_j=(c_0/c_2)(|m|-1)$, and $x^*_m$ is arbitrary or free, determined only by the boundary condition. 

Linear stability analysis (Supplementary information) shows that
chief-clan structures with a single chief are stable, which was tested
for strand lengths up to 20. It also shows that most of the chief-clan
structure having two or four chiefs are stable. The probability to find
a Jacobian with complex eigenvalues increases rapidly with the increase
of chief length for the case of autocatalytic systems ($c_1=0$). On the
other hand, the eigenvalues are always real and non-positive for the
case of random systems ($c_2=0$). Based on this analysis, we believe
that chief-clan structures are almost always locally stable for
autocatalytic systems. Yet, we
will see below that only very few chief-clan structures out of the vast
number of possible structures are selected by the dynamics.

Now let us consider what kind of chief-clan structure is expected to appear from the pool of monomers if ligation is mostly autocatalytic, but there exists a small rate of random ligation ($c_1\bar{x}\ll c_2$).
The constant solution with unconstrained chiefs drives the selection of chiefs with more regular sequence than irregular one as follows. 

Once a (short or intermediate) chief is created by random ligation, the
constant solution [\eqref{constant} and \eqref{eq_chief}]
equilibrates the population of its clan members toward the constant
value $\bar x$ and the rest of the material is accumulated into the
chief.  Thus, the population is ``inverted'' and the next random
ligation likely takes place between the chiefs.  The probability of
ligation between other species than chiefs decreases rapidly.  This
creates a regularity of the chiefs; e.g. an intermediate chief 01
creates another intermediate chief 0101, which then creates 01010101.
Moreover, 0101 can ligate with 0101 at both ends to create 01010101,
whereas an irregular chief demands its intermediate substrands to ligate
in correct order.  Thus, regular chiefs such as 010101... or
00110011... are likely selected because of the constant solution.  This
selection spontaneously appears out of autocatalytic ligations in the
systems.


\begin{figure}
 \begin{center}
  \includegraphics[width=\columnwidth]{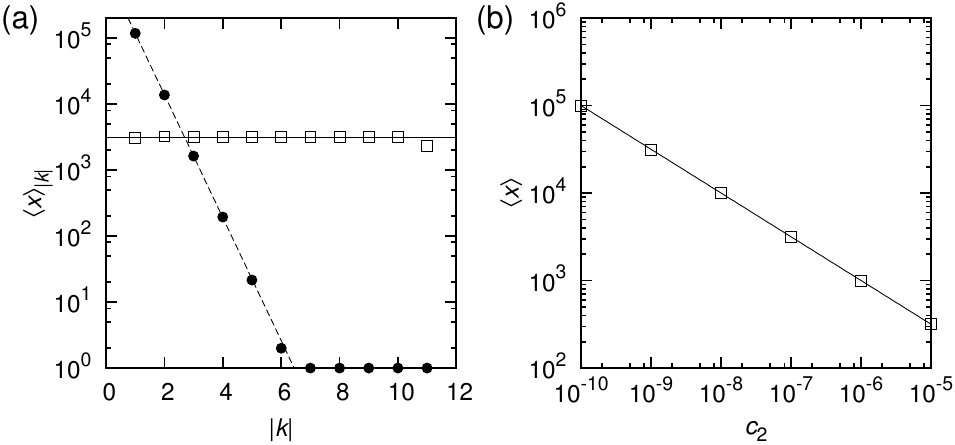} \caption{Confirmation
  of the exponential solution \eqref{exponential} for the random systems
  and the constant solution \eqref{constant} for the autocatalytic
  systems. (a) The mean value of $x_k$ for species with the same $|k|$
  is plotted against $|k|$. Solid circles represent the results for a
  random system ($c_0=1,c_1=1\times 10^{-6},c_2=0$). The dashed line is
  $(c_0/c_1)\exp(-b|k|)$ with $b\simeq 0.5$. Open squares represent the
  results for autocatalytic systems ($c_0=1,c_1=0,c_2=1\times 10^{-7}$),
  when the bootlace structure [Fig.~\ref{structure}(a)] was given as an
  initial condition. The solid line is the constant solution,
  $\bar{x}=\sqrt{c_0/c_2}$. The longest species is the chief ($|k|$ =
  11), whose concentration is unconstrained. (b) The mean value of the
  concentration against $c_2$ with the fixed value of $c_0=1,c_1=0$. The
  solid line is $\bar{x}=\sqrt{c_0/c_2}$.}  \label{dist}
 \end{center}
\end{figure}

To confirm the behaviors of the system considered above, we sample
stochastic trajectories from numerical simulation of the microscopic
reactions \eqref{eq_decomposition} - \eqref{eq_catalyzed}, using the
exact stochastic simulation algorithm by Gillespie~\cite{Gillespie1976}.
The system is assumed to be well stirred and its volume is set to
1. Simulations are started in the state of 200,000 monomers of 0 and
200,000 monomers of 1.  The chemical reaction constants are set to
$c_0=1$, $c_1=1\times 10^{-10}$, and $c_2=1\times 10^{-7}$, unless
otherwise noted.  With $c_0=1$ the unit of time $t$ is the relaxation
time of decomposition.

Figure~\ref{dist}(a) shows how $x_k$ obtained by simulation depends on
the length $|k|$.  The exponential solution is always observed in purely
random systems [$c_2=0$, solid circles in Fig.~\ref{dist}(a)] as
predicted. The constant solution, $\bar{x}=\sqrt{c_0/c_2}$, on the other
hand, is always observed in purely autocatalytic systems [$c_1=0$,
Fig.~\ref{dist}(a)].  The concentration of the chiefs is also confirmed
to be free from $\bar{x}$, as a point deviated from $\bar{x}$ in
Fig.~\ref{dist}(a).  The dependence of $\bar{x}$ on $c_2$ is shown in
Fig.~\ref{dist}(b).  The theory ($\bar{x}=\sqrt{c_0/c_2}$, solid line)
predicts the exact values obtained by the simulation.

Now, let us describe the simulation results starting from the pool of monomers.
In purely random systems ($c_2=0$), there is no selection of specific sequence structures: 
species decrease in concentration exponentially proportional to their strand length $|k|$, but their concentration $x_k$ is independent of the particular sequence information $k\in\mathcal A^*$.

In contrast, in autocatalytic systems with small rate of random ligation ($c_1\bar{x}\ll c_2$) only very few distinctive and highly ordered population structures and sequence patterns are selected out of a huge number of possibilities.
Different populations $x$ and $x'$ can be compared by their cosine distance $d(x,x')=1-\cos(x,x')$.
We classify the populations obtained by simulation with a single linkage hierarchical cluster algorithm (HCA) \cite{Johnson1967}
based on $d$.
Clustering is stopped when 80\% of the populations is grouped into clusters.

\begin{figure}
 \begin{center}
	\includegraphics[width=.97\columnwidth]{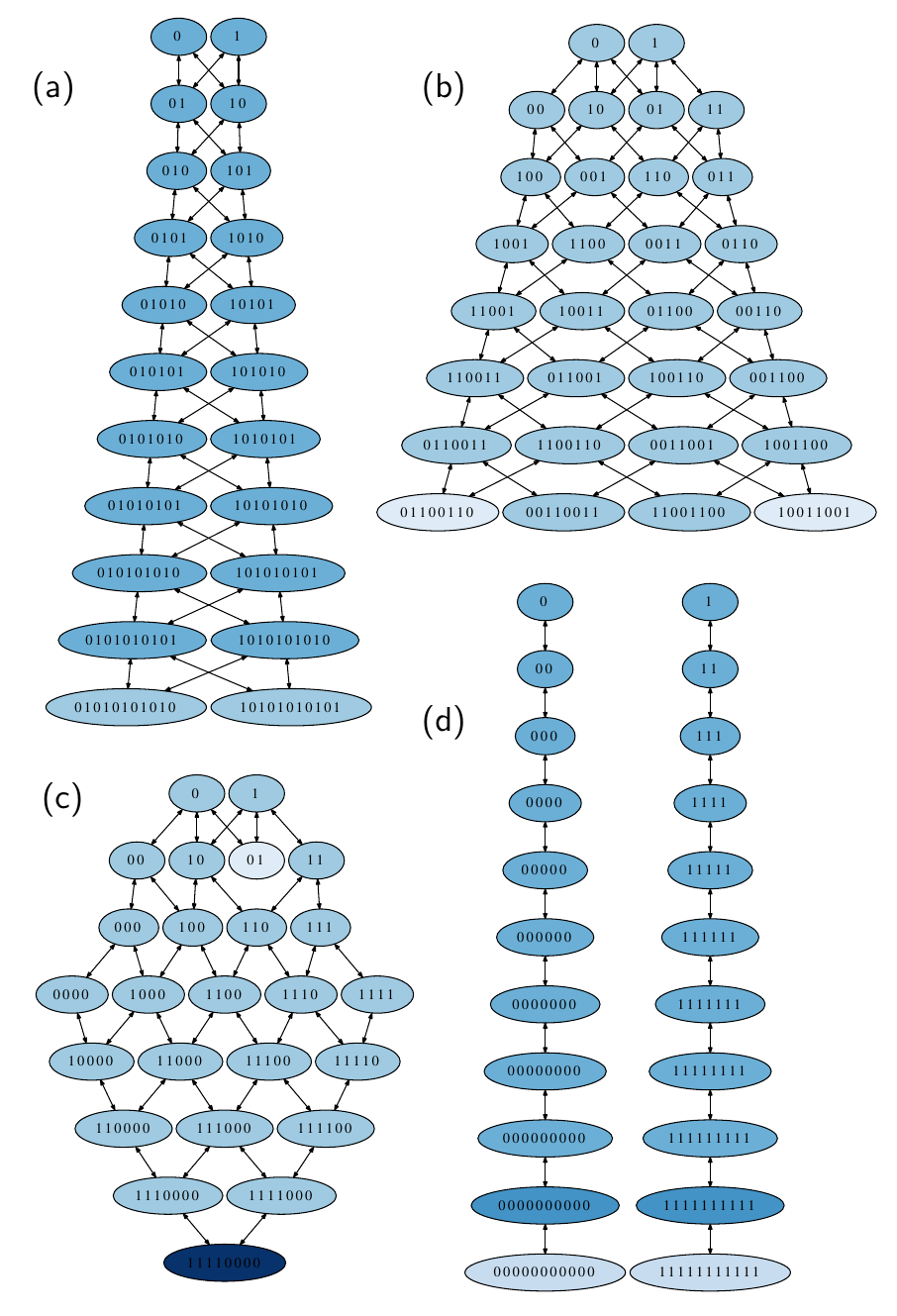}
	\caption{Four most frequent structures seen in the simulation at
	$t=100$. Each structure is the result of average over 109-2034
	structures, named: (a) bootlace, (b) pinecone1, (c)
	pinecone2, and (d) two towers. The darkness shows the
	concentration of each molecules, where the darker the larger the
	concentration. Species having less concentration than
	10\% of the constant solution in \eqref{constant} are omitted.}
	\label{structure}
 \end{center}
\end{figure}

HCA shows that only four structures occupy about 70\% of all populations produced by simulation of autocatalytic systems.
We call them, from the most frequent one, bootlace, pinecone1, pinecone2, and two towers (Fig.~\ref{structure}).
Their probabilities of appearance at $t=100$ are about 45\%, 19\%, 5\%, and 2\%, respectively.  
Note that populations with these distinctively regular structures always have a Jacobian with non-positive real eigenvalues. 

It can be seen that the selected populations have only few chiefs, where deviation from the constant solution is possible (lighter or darker species in Fig.~\ref{structure}).
Though fluctuations sometimes create shorter chiefs, their concentration is always much lower than $\bar{x}$, and they usually disappear quickly.


To confirm the mechanisms of this selection of the structures, a typical process of its appearance is shown in Fig.~\ref{process}.
After spontaneous formation by random ligation, an intermediate chief (10) accumulates material through autocatalysis [Fig.~\ref{process}(a)] in order to satisfy the constant solution [Eq.~\eqref{constant} and \eqref{eq_chief}].
Progressively longer chiefs are formed primarily by random ligation of previous chiefs, since
they gather the majority of material [Fig.~\ref{process}(b)].
Once created, chiefs decompose to produce their clan members [Fig.~\ref{process}(b)$\to$(c) and (d)$\to$(e)].
This continues until chief concentrations become significantly lower than $\bar x$ at which point there is not enough material to create and stabilize longer chiefs.
The overlap of clans introduces resource competition.
Survival against this competition is biased, since the material flow rate from the existing species to new chiefs is not equal among the clans:
a chief with more and larger populated exclusive clan members is expected to collect material faster and is therefore more likely to survive competition (Supplementary Information).
Thus a short chief (100) in Fig.~\ref{process}(f) is likely to disappear. 

The described selection mechanism is absent in purely random ligation
because the exponential equilibrium solution [Eq.~\eqref{exponential}] does not allow the ``inverted'' population structure where material is accumulated in chiefs and there is thus no driving force to push the system into creating longer species.  Thus the selection only arises out of autocatalysis. 

We can estimate the likelihood of forming certain clan structures and sequence patterns from monomer pools of 0 and 1 by assuming relatively seldom random ligation of chiefs, subsequent equilibration, and competition among intermediate chiefs.
The details of this calculation are given in the supplementary information.
The calculated likelihoods are quantitatively in good agreement with the structures observed in simulation.
For varying system sizes, these calculations further suggest that the two towers structure becomes less likely with increasing system size, whereas system size shows little impact on the probabilities to obtain other regular structures.

We now discuss the stability of the four selected structures against fluctuation, quantified by the temporal change of cosine similarity, $I(t)=d(x(0),x(t))$, between the structure at time $t$ and the initial one.
Simulations are started from already stabilized structures.

\begin{figure}
 \begin{center}
	\includegraphics[width=\columnwidth]{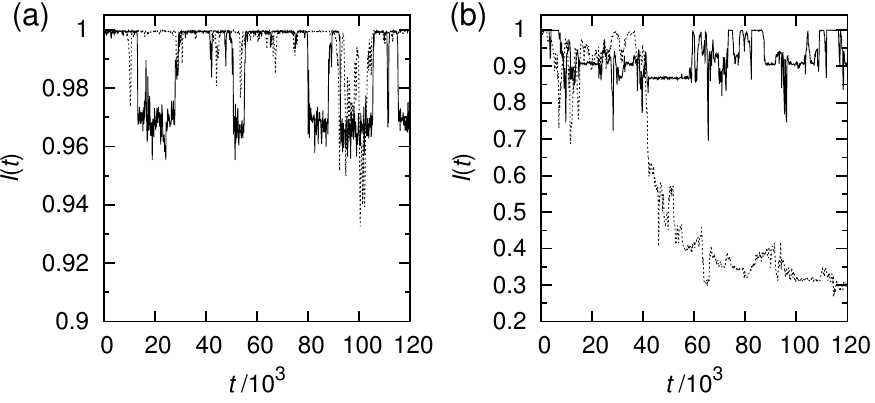}
  \caption{Fluctuations seen in the temporal change of cosine similarity
  between the structure at $t=0$ and the one at $t$. (a) bootlace, solid
 line and two towers, dotted line. (b) pinecone1, solid line and
 pinecone2, dotted line. Note the difference of the scale of $y$-axis.}
 \label{fluctuation}
 \end{center}
\end{figure}

Bootlace and two towers are mostly stable and exhibit only small fluctuations [Fig.~\ref{fluctuation}(a)] compared with those of pinecones [Fig.~\ref{fluctuation}(b)].
Note the different $y$-axis scales.
The bootlace structure fluctuates between two stable states, where the structure suddenly destabilizes after a period of stability, and switches into another stable structure.
For bootlace, this is likely attributed to the competition between the chiefs 10101... and 01010....
Another such punctuated equilibrium is observed in pinecone1 [Fig.~\ref{fluctuation}(b)], where it originates from the switching among the four chiefs 00110011..., 01100110..., 11001100..., and 10011001....

Even a transition from a structure to a completely different one can occur.
Pinecone2 [Fig.~\ref{structure}(c)] is metastable under this boundary condition, and after a long time suddenly transitions into a different structure [sudden drop of dotted line in Fig.~\ref{fluctuation}(b)].
This transition is caused by the inverted population of the chief [11110000 in Fig.~\ref{structure}(c)].
A random ligation creates more stable chiefs from this chief, and the new chief-clan structure relaxes the inverted population.
This transition suggests that the metastability is determined by the boundary condition, that is, how much material flows into chiefs through their clan members.

In conclusion, we find that very limited combinations of cooperative species are selected out of a huge number of possibilities in a system of autocatalytic bit matching polymers.
Particular polymer sequence with repetitive patterns are selected and polymers with these patterns together generate their associated macroscopic cooperative population structure.
The selected populations have a ``chief-clan'' structure, where all the clan members but chiefs have the same concentration.
The analytic solutions for the reaction kinetics are quantitatively confirmed by microscopic stochastic simulations.
The stability of the structures depends on boundary conditions.
Under the boundary condition used, four emergent structures are found:
three of them are stable, while another one is metastable and successively transitions into a different structure.

Without autocatalysis, no such selection takes place and species appear with concentration exponentially decreasing with sequence length.
Selection and cooperation of structures arise out of autocatalysis and symmetries in the reaction network and do not require a predefined fitness function to induce selection pressure.

We suspect that many of the reported findings also appear in more detailed model systems with strand complementarity, sequence overhang, mismatches~\cite{Eigen1971}, hybridization behavior~\cite{Fernando2007,Fellermann2011}, and convection or diffusion limited reactions~\cite{Obermayer2011,Walker2012}.
Thus, our analysis might provide clues about expected early information-metabolism patterns in the origin of life.

\end{document}